# Privacy-Preserving Methods for Vertically Partitioned Incomplete Data

Yi Deng, PhD[1], Xiaoqian Jiang, PhD[2], Qi Long, PhD[3]
[1]Emory, Atlanta, GA, USA; [2]University of Texas Health Science Center, Houston, TX, USA; [3]University of Pennsylvania, Philadelphia, PA, USA

**Abstract**

*Distributed health data networks that use information from multiple sources have drawn substantial interest in recent years. However, missing data are prevalent in such networks and present significant analytical challenges. The current state-of-the-art methods for handling missing data require pooling data into a central repository before analysis, which may not be possible in a distributed health data network. In this paper, we propose a privacy-preserving distributed analysis framework for handling missing data when data are vertically partitioned. In this framework, each institution with a particular data source utilizes the local private data to calculate necessary intermediate aggregated statistics, which are then shared to build a global model for handling missing data. To evaluate our proposed methods, we conduct simulation studies that clearly demonstrate that the proposed privacy-preserving methods perform as well as the methods using the pooled data and outperform several naïve methods. We further illustrate the proposed methods through the analysis of a real dataset. The proposed framework for handling vertically partitioned incomplete data is substantially more privacy-preserving than methods that require pooling of the data, since no individual-level data are shared, which can lower hurdles for collaboration across multiple institutions and build stronger public trust.*

**Introduction**

The last decade has witnessed tremendous growth of biomedical data that is routinely collected from various sources. For instance, hospitals deposit electronic health records (EHRs) into medical databases, genome sequencing centers collect human genome data, and wearable devices contribute to another major source of mobile health data. The US federal government and other nonprofit organizations have been vastly acquiring health-care knowledge, including data from clinical units and information about patients from insurance companies. Large data networks such as PCORnets have been established to make it faster, easier, and less costly to conduct clinical research than is now possible by harnessing the power of large amounts of health data across multiple institutions [1].

There is an ever-increasing need to develop statistical methods for analyzing data within such networks. However, the processes of integrating data pose real privacy issues: data that are of restricted sensitivity may become highly sensitive after being integrated (pooled). For example, the clinical diagnosis data with patients' demographics in the combined dataset are highly sensitive, and sharing them across networks (institutions) may lead to a high risk of information disclosure. There is a growing body of literature showing that a unique combination of patient characteristics [2,3], diagnose pattern [4], hospital visit trials [5,6], and even the length and frequency of visits might lead to identification [7]. Research is rapidly progressing to propose novel privacy-preserving approaches that can overcome such privacy issues. Vaidya and Clifton [8] adopt a conceptually simple definition of "privacy": a collaborating institution should learn nothing from any other institution's data. The idea is to minimize the information sharing unless necessary to reduce the potential attack surface as much as possible. In order to protect privacy and reduce disclosure risk, it is common for institutions to manipulate (e.g., perturb or coarsen) the data prior to integrating [9]. However, such perturbed (or synthetic) data bring reduced utilities and imprecise conclusions as well [10]. We are aware of differential privacy techniques [11], but missing data imputation algorithms are sensitive to the noise (and sometimes require multiple fill-ins) and might lead to a very limited utility. In this paper, we are proposing a practical approach to balance the utility and privacy tradeoff, which is to compute and share only the summary statistics instead of the sensitive patient-level data. Within the statistics literature, most attention has been drawn into the case of horizontally partitioned data [12,13]. In this paper, we investigate a more common case that data are vertically partitioned.

Vertically partitioned data refer to the data from different institutions that have mutually exclusive characteristics, features, or variables for the same population. This is commonly present in real collaborations among different types of data providers. For instance, local and federal agencies, hospitals, and private corporations with different information about the same population can work together to develop comprehensive quantitative models to produce meaningful results. A variety of privacy-preserving methods have been proposed to address statistical tasks, including linear regression, [14,15], and logistic regression [16]. Other works on mining vertically partitioned data include linear discriminant analysis [17][7], association rule mining [18], support vector machine [19], naïve Bayes [8] and k-means [20]. From a statistical perspective, some of these techniques proposed by computer scientists are incomplete in a way that, for example, coefficient estimates are provided, while standard errors and other essential statistics for inferences are ignored. Those neglect statistics are of decisive importance in some biomedical research, especially association studies. Karr *et al.* [14] propose a protocol for model diagnostics via secure matrix multiplications. However, their method requires high communication costs and heavy computation when the number of institutions is large, which is not scalable.

Although the developments of privacy-preserving alternatives of the standard statistical learning techniques are extensive, research on how to deal with missing values for such vertically partitioned data is absent. In addition, with the prevalence of distributed networks and an increasing number of institutions participating, investigators experience missing values more frequently. These two factors motivate us to propose privacy-preserving methods for incomplete, vertically partitioned data. Specifically, assuming the data follow a univariate missing data pattern, we propose two privacy-preserving approaches that couple distributed models (linear regression and logistic regression) with an inverse probability weighting (IPW) technique and a multiple imputation (MI) technique, respectively. Our privacy-preserving IPW for vertically partitioned data (PPIPW-V) first builds a distributed logistic regression model on the probability of observing a complete case, without disclosing individual-level data. Then we calculate the weights as the inverse of the estimated probabilities and fit a weighted distributed linear regression model assuming our original analysis model of interest is the multiple linear regression. PPIPW-V can be easily extended to the case of logistic regression of a binary outcome variable on independent variables that are collected by different institutions. MI methods for handling missing data are popular and are shown to perform well in both literature and practice [21–23]. We propose a privacy-preserving MI approach for vertically partitioned data (PPMI-V) assuming the response variable is fully observed while one independent variable may be missing partially on a subset of records. Utilizing the technique of multiple imputations by chained equations (MICE) [24], we can extend PPMI-V to be applicable to data that have general missing data patterns. We offer guidance and suggestions to calculate standard errors for both PPIPW-V and PPMI-V through bootstrap resampling.

The remainder of this paper is organized as follows. In the Methods section, we formulate and describe the settings and notation of missing data that are vertically partitioned. Then, we formally develop the proposed methods, namely PPIPW-V and PPMI-V. In the Numerical Experiments section, we demonstrate that the proposed methods for vertically partitioned incomplete data perform as well as the existing IPW and MI methods that use pooled data. We also provide some practical recommendations for applications. Furthermore, we generate "synthetic" incomplete data from the Georgia Coverdell Acute Stroke Registry (GCASR) data to mimic the case that data are vertically partitioned. PPIPW-V and PPMI-V, as well as other methods, are applied to the synthetic data for comparisons. This empirical study demonstrates the effectiveness of our proposed methods in real data applications.

**Methods**

We investigate vertically partitioned data and propose privacy-preserving approaches for handling missing data in such settings. To fix ideas, consider a linear regression model $Y = \mathbf{X}\theta + \varepsilon$. The objective of the regression analysis is to estimate regression coefficients $\theta$, when the covariate $\mathbf{X}$ is subject to missing values. For vertically partitioned data from a distributed environment with $K$ institutions which are referred to as sites in the mathematical formula, we let $\mathbf{X} = (\mathbf{X}^{site_1}, \ldots, \mathbf{X}^{site_K})$, where $\mathbf{X}^{site_k}$ is a set of covariates collected from the $k$-th institution for the same group of patients (of sample size $= n$). Of note, the more general setting where some covariates can be shared or are available across institutions is equivalent to a vertically partitioned setting where the set of shared covariates reside in one single site for which our proposed methods can be applied. We assume that the outcome variable $Y$ is accessible to all institutions. Let the total number of covariates be $p$ and denote the number of covariates in the $k$-th institution by $p_k$ and $\sum_{k=1}^{K} p_k = p$. Without loss of generality, we assume $\mathbf{X}^{site_1}$ has a vector of all 1's to include an "intercept" term in the regression model, i.e., $\mathbf{X}^{site_1} = (1, X_1, \ldots, X_{p_1})$. To illustrate our ideas, we consider a univariate missing data pattern where only $X_1$ (from $\mathbf{X}^{site_1}$) has missing values and the other variables are fully observed. It is straightforward to extend our methods to general missing data patterns.

*Privacy-preserving inverse probability weighting for vertically partitioned incomplete data (PPIPW-V)*

To apply IPW on vertically partitioned incomplete data, we need to develop a distributed logistic regression model for the weights (Stage 1) and a distributed linear regression model for the weighted subjects (Stage 2).

In Stage 1, we fit a non-response model to predict the weight for each patient. Let us denote the predictors of the probability of observing $X_1$ by $\mathbf{Z} = (1, Y, X_2, \ldots, X_p)$. For notational convenience, let $\mathbf{Z} = (\mathbf{Z}^{site_1}, \ldots, \mathbf{Z}^{site_K})$, where $\mathbf{Z}^{site_1} = (1, Y, X_2, \ldots, X_{p_1})$ and $\mathbf{Z}^{site_k} \equiv \mathbf{X}^{site_k}$, for $k \geq 2$.

Unlike horizontally partitioned data where $\mathbf{Z}^T \mathbf{W}^{old} \mathbf{Z} = \sum_{k=1}^{K} (\mathbf{Z}^{site_k})^T \mathbf{W}^{old} \mathbf{Z}^{site_k}$, the distributed logistic regression for vertically partitioned data is quite different. We adopt the approach proposed by Li *et al.* [16] that optimizes the logistic regression model by dual optimization. The original maximization of the log-likelihood of a primal problem is replaced with the minimization of the dual form log-likelihood, which guarantees the same optimum. After the transformation, the linear decomposition becomes feasible for dual optimization. We use a response indicator $s_i$, taking value 1 if variables for individual $i$ are fully observed and -1 otherwise, to better represent the primal form of the log-likelihood function. The details of Stage 1 of PPIPW-V can be found in the appendix. Figure 1 summarizes those steps in Stage 1:

- Step 1: using the most recent dual parameter estimates, each site updates the primal parameters for its own features (of size $p_k \times 1$) and then calculate two site-level aggregated statistics (each of size $n \times 1$). Since the calculation in an individual site does not depend on other sites, Step 1 can be done in parallel across all sites.
- Step 2: each site sends two site-level aggregated statistics to the master site. Since the statistics are aggregated, no patient-level features are shared.
- Step 3: after receiving all site-level aggregated statistics, the master site sums them up to get overall aggregated statistics and only use them to update the dual parameters.
- Step 4: the master site sends the updated dual parameter estimates (of size $n \times 1$) to each site.

Steps 1-4 are repeated until the parameter estimates converge. One final step in Stage 1 of PPIPW-V is that the master site calculates the predicted propensity score (weight) for all patients.

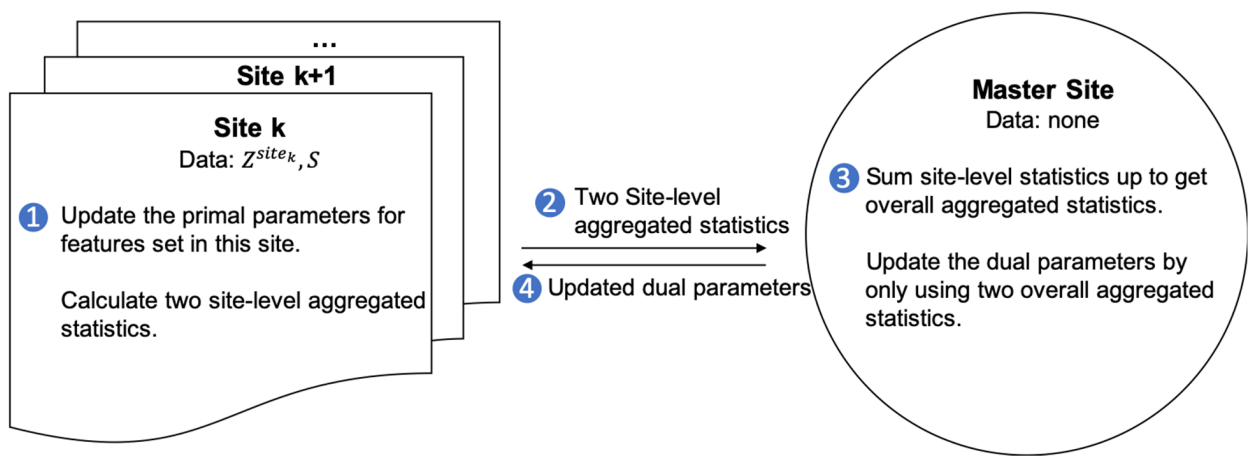

**Figure 1**: Flow chart of fitting distributed logistic regression (Stage 1) in PPIPW-V.

In Stage 2, we establish a weighted distributed linear regression model in this vertically partitioned setting, with the weights obtained from Stage 1. The objective function can be written in the matrix form: $F(\theta) = (Y - \mathbf{X}\theta)^T \mathbf{V}(Y - \mathbf{X}\theta)$, where $\mathbf{V}$ is the diagonal matrix of weights. In the case of IPW, $\mathbf{V} = diag(\{r_i \widehat{w}_i\}_{i=1}^n)$. The quadratic programming problem of minimizing $F(\theta)$ can be solved by a derivative-free modified Powell's algorithm proposed by Sanil *et al.* [15], which can be implemented in the vertically partitioned health data networks without sharing patients' features. Of note, the implementation of the distributed Powell's algorithm (see details in the appendix) does not require the data to be pooled in one place and only aggregated summary statistics are shared between sites iteratively to make the algorithm converge.

Powell [25] showed that if $F(\theta)$ is a quadratic function, the proposed algorithm would yield the exact minimizer $\widehat{\theta} = arg\ min_\theta F(\theta)$. The asymptotic distribution of the weighted parametric estimate of $\theta$ is derived by Wang *et al.* [26]. Their rigorous estimated covariance matrix is sophisticated and difficult to calculate in the distributed environments. We use bootstrap to approximate the standard error of $\hat{\theta}$, for practical considerations. To generate bootstrap data $(Y, \mathbf{X})^b$ repeatedly using vertically distributed data, we sample the indices $(1, \ldots, n)$ with replacement and share them to all institutions, which prepare (arrange) the dataset accordingly.

*Privacy-preserving multiple imputations for vertically partitioned incomplete data (PPMI-V)*

Suppose the missing variable $X_1$ is continuous and follows a normal distribution given $Y$ and other covariates. That is, $X_1 = \mathbf{Z}\alpha + \varepsilon$, where $\varepsilon \sim_{iid} N(0, \sigma^2)$. We can use complete cases to estimate parameter $\alpha$. The objective function (sum of squared residuals) is $F(\alpha) = (X_1 - \mathbf{Z}\alpha)^T \mathbf{V}(X_1 - \mathbf{Z}\alpha)$, with $\mathbf{V} = diag(\{r_i\}_{i=1}^n)$. Since $F(\alpha)$ is in a quadratic form, derivative-free modified Powell's algorithm can be directly applied to get the least square estimator $\hat{\alpha} = arg\,min_\alpha F(\alpha)$. The intermediate value $\delta$ can be calculated using summary statistics from each institution, leading to obtaining $\hat{\alpha}$ with confidentiality. We can then take advantage of the bootstrap technique to estimate the variance of $\hat{\alpha}$, denoted by $\widehat{\mathbf{V}}_\alpha$. The multiple imputation method first draws a value $(\hat{\alpha}^{(m)}, \hat{\sigma}^{(m)2})$ from the posterior distribution of $(\alpha, \sigma^2)$, where $\hat{\alpha}^{(m)}$ is drawn from a multivariate normal distribution with mean $\hat{\alpha}$ and variance matrix $\widehat{\mathbf{V}}_\alpha$, and $\hat{\sigma}^{(m)2}$ is drawn from $\sum_{i=1}^n r_i(x_{1,i} - \sum_{k=1}^K (\mathbf{z}_i^{site_k})^T \hat{\alpha}^{(m)site_k})^2 / \chi^2_{(\sum r_i - p - 1)}$. For individuals $i$ missing $X_1$, given observed $\mathbf{z}_i$, $X_{1,i}$ is then drawn from a $N(\mathbf{z}_i^T \hat{\alpha}^{(m)}, \hat{\sigma}^{(m)2})$, where $\mathbf{z}_i^T \hat{\alpha}^{(m)} = \sum_{k=1}^K (\mathbf{z}_i^{site_k})^T \hat{\alpha}^{(m)site_k}$ is linearly decomposable. We repeat the above procedure $M$ times and create $M$ multiply imputed datasets. Each dataset is then processed by the aforementioned Modified Powell's algorithm, resulting in $M$ estimates $\{\hat{\theta}^{(m)}\}_{m=1}^M$ of the parameters of interest $\theta$. The final estimate can be obtained by combining $\{\hat{\theta}^{(m)}\}_{m=1}^M$ using Rubin's rules [27].

**Numerical Experiments**

*Simulated Data*

We evaluate in this section the performance of the proposed methods. Suppose our vertically partitioned data $\mathbf{X}$ consist $p = 6$ independent variables from $K = 3$ institutions. We assume each institution possesses two independent variables and has access to the outcome variable $Y$. We are interested in multiple linear regression:

$$Y = \mathbf{X}\theta + \varepsilon$$
$$= (1, \overbrace{X_1, X_2}^{site_1}, \overbrace{X_3, X_4}^{site_2}, \overbrace{X_5, X_6}^{site_3}) \times (\theta_0, \theta_1, \theta_2, \theta_3, \theta_4, \theta_5, \theta_6)^T + \varepsilon,$$

when $X_1$ has missing values.

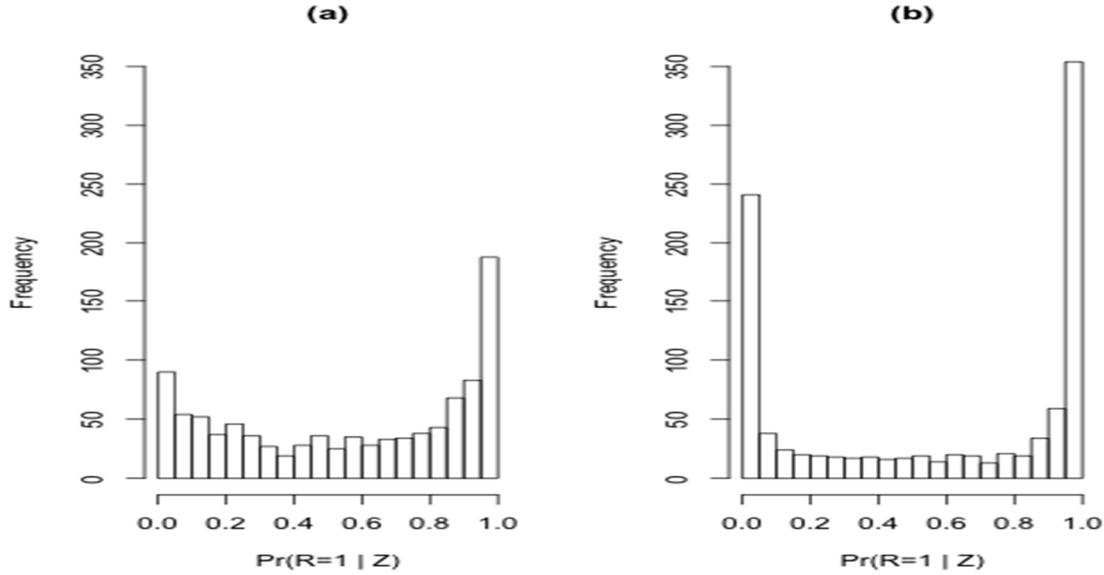

**Figure 2**: Histograms of the probabilities of observing a complete case in (a) Scenario 1 and (b) Scenario 2, using 1000 subjects.

We generate $X_2, \ldots, X_6$ independently from the uniform distribution on (-1, 1). We then generate $X_1$ from a normal distribution with mean $\sum_{j=2}^{5} X_j/\sqrt{5}$ and variance 1. The continuous variable $Y$ is generated from $\mathcal{N}(\mathbf{X}\theta, \sigma^2)$, where $\theta = (1,1,0,1,0,1,0)^T$ and $\sigma^2 = 1$. IPW methods are known to suffer from the issue of extreme weights. To evaluate the impact of extreme weights in the vertically partitioned setting, we consider two scenarios for the selection probabilities as follows:

Scenario 1.  $Pr(R = 1|Y, \mathbf{X}) = \{1 + \exp(-1.6 + Y + X_3 + X_5)\}^{-1}$

Scenario 2.  $Pr(R = 1|Y, \mathbf{X}) = \{1 + \exp(3 + 2Y + 2X_3 + 2X_5)\}^{-1}$.

Since the missingness of $X_1$ does not depend on itself, the data are missing at random (MAR). Figure 2 presents examples of the frequency plots of the probabilities in the two scenarios using n=1000 subjects. While both scenarios lead to about the same 42% of subjects missing $X_1$, Scenario 1 provides relatively stable weights compared to Scenario 2, where weights are defined as the inverse of the probabilities.

We compare our proposed methods with standard approaches including inverse probability weighting methods and multiple imputation. Specifically, we consider the following seven methods:

- Gold standard (GS): the analysis on n subjects with underlying true values for missing data
- Complete-cases (CC): the analysis on fully observed subjects using a distributed linear regression
- IPW-pooled: the standard IPW approach on pooled data throughout the whole process of estimating weights and fitting weighted linear regression
- PPIPW-V: the proposed privacy-preserving IPW for vertically partitioned data;
- MI-naive: each institution imputes the missing data using their own data, following by applying a distributed linear regression for the analysis model
- MI-pooled: the standard MI approach on pooled data throughout the whole process of

predicting missing values (M=100 times) and fitting a standard linear regression.
- PPMI-V: the proposed privacy-preserving MI with M=100 imputed datasets for vertically partitioned incomplete data

**Table 1:** Simulation results for estimating $\theta_1 = \theta_3 = \theta_5 = 1$ based on 1000 Monte Carlo replications for Scenario 1 with sample size n = 200 or 1000. RBias, mean relative bias; SE, mean, standard error; SD, Monte Carlo standard deviation; MSE, mean square error; CR, the coverage rate of 95% confidence interval.

| N | Method | $\hat{\theta}_1$ | | | | | $\hat{\theta}_3$ | | | | | $\hat{\theta}_5$ | | | | |
|---|---|---|---|---|---|---|---|---|---|---|---|---|---|---|---|---|
| | | RBias (%) | SE | SD | MSE | CR (%) | RBias (%) | SE | SD | MSE | CR (%) | RBias (%) | SE | SD | MSE | CR (%) |
| 200 | GS | 0.058 | 0.065 | 0.066 | 0.004 | 95.1 | 0.259 | 0.127 | 0.131 | 0.017 | 93.2 | -0.260 | 0.127 | 0.127 | 0.016 | 96.1 |
| | CC | -9.427 | 0.088 | 0.087 | 0.016 | 80.5 | -19.702 | 0.169 | 0.174 | 0.069 | 77.9 | -20.409 | 0.170 | 0.163 | 0.068 | 76.9 |
| | IPW-pooled | -2.964 | 0.113 | 0.139 | 0.020 | 88.3 | -3.832 | 0.220 | 0.264 | 0.071 | 88.6 | -6.445 | 0.220 | 0.265 | 0.074 | 88.1 |
| | PPIPW-V | -3.293 | 0.108 | 0.129 | 0.018 | 89.0 | -6.454 | 0.209 | 0.247 | 0.065 | 88.5 | -8.787 | 0.210 | 0.243 | 0.067 | 87.1 |
| | MI-naïve | -3.227 | 0.083 | 0.080 | 0.007 | 94.3 | -9.313 | 0.158 | 0.139 | 0.028 | 93.2 | -9.740 | 0.158 | 0.140 | 0.029 | 93.7 |
| | MI-pooled | -1.897 | 0.079 | 0.078 | 0.006 | 95.0 | 0.847 | 0.161 | 0.156 | 0.024 | 95.8 | 0.583 | 0.161 | 0.159 | 0.025 | 95.5 |
| | PPMI-V | -1.620 | 0.074 | 0.079 | 0.006 | 92.9 | 1.410 | 0.167 | 0.158 | 0.025 | 96.0 | 1.265 | 0.166 | 0.160 | 0.026 | 95.5 |
| 1000 | GS | -0.096 | 0.029 | 0.028 | 0.001 | 95.0 | 0.128 | 0.056 | 0.056 | 0.003 | 94.8 | -0.117 | 0.056 | 0.056 | 0.003 | 94.9 |
| | CC | -9.451 | 0.039 | 0.039 | 0.010 | 31.1 | -19.975 | 0.075 | 0.074 | 0.045 | 23.9 | -20.052 | 0.075 | 0.076 | 0.046 | 25.0 |
| | IPW-pooled | -0.447 | 0.065 | 0.078 | 0.006 | 89.6 | -1.318 | 0.122 | 0.142 | 0.020 | 89.6 | -1.740 | 0.122 | 0.142 | 0.021 | 89.8 |
| | PPIPW-V | -0.467 | 0.064 | 0.077 | 0.006 | 89.6 | -1.867 | 0.121 | 0.140 | 0.020 | 88.0 | -2.252 | 0.121 | 0.141 | 0.020 | 89.0 |
| | MI-naïve | -2.419 | 0.035 | 0.034 | 0.002 | 90.9 | -9.900 | 0.069 | 0.061 | 0.013 | 72.2 | -9.994 | 0.069 | 0.063 | 0.014 | 71.4 |
| | MI-pooled | -0.257 | 0.034 | 0.033 | 0.001 | 95.4 | 0.392 | 0.070 | 0.069 | 0.005 | 95.4 | 0.135 | 0.071 | 0.071 | 0.005 | 94.5 |
| | PPMI-V | -0.209 | 0.032 | 0.033 | 0.001 | 93.8 | 0.126 | 0.072 | 0.069 | 0.005 | 96.1 | -0.086 | 0.072 | 0.070 | 0.005 | 95.7 |

Table 1 shows the simulation results from Scenario 1. The data are missing at random, and the missingness mechanism depends on the outcome variable $Y$. We first illustrate the performances of standard IPW and MI using pooled data (i.e., IPW-pooled and MI-pooled). IPW-pooled generates unbiased estimators while CC provides a large bias. This finding is more clear when $n = 1000$ that the relative bias of IPW-pooled is negligible. This means that by applying the weights, IPW-pooled can, as expected, reduce the bias compared to CC, which also uses the complete cases only. However, the results show that the estimators of IPW-pooled, even though unbiased, are still under covered by 95% confidence intervals. Of note, IPW-pooled also possesses a much larger SE than others. The MI-pooled method, on the contrary, yields an unbiased estimator with a relatively small SE. It also has good coverages that are all close to the 95% level. In terms of vertically partitioned incomplete data, a naïve way of handling missing values is to replace them with predicted ones using other covariates from the same institution. We denote this method by MI-naïve. The imputation models in MI-naïve are improper since the missing variable may actually depend on covariates of other institutions. Thus, ignoring those covariates will generally lead to biased estimators. As shown in Table 1, MI-naïve has large biases and serious low coverage rates, which correspond to the fact it is inconsistent. PPIPW-V and PPMI-V inherit the property of standard IPW and MI using pooled data, respectively. They perform in a similar way to their corresponding versions of non-distributed methods. In other words, PPIPW-V gives unbiased results but large SEs, as IPW-pool; PPMI-V provides unbiased estimators with small SEs, as MI-pooled. This interesting finding confirms that our proposed privacy-preserving methods on vertically partitioned data work as well as the methods used on pooled data. They offer solutions to the missing data problem in distributed data networks by providing meaningful results without individual-level data.

Table 2 displays the simulation results from Scenario 2, which has unstable weights. A considerable number of probabilities of observing a complete case are very close to 0, leading to extremely large weights. It is possible that in practice, the logistic model of missingness yields very large weights for some individuals with moderate weights, due to lack of fit. On the one hand, the results show that both IPW-pooled and PPIPW-V have biased estimators that give rise to large MSE. MI-pooled and PPMI-V, on the other hand, do not require the specification of the missingness model and perform well in this scenario.

**Table 2**: Simulation results for estimating $\theta_1 = \theta_3 = \theta_5 = 1$ based on 1000 Monte Carlo replications for Scenario 2 with sample size n = 200 or 1000. RBias, mean relative bias; SE, mean standard error; SD, Monte Carlo standard deviation; MSE, mean square error; CR, the coverage rate of 95% confidence interval.

| N | Method | $\hat{\theta}_1$ | | | | | $\hat{\theta}_3$ | | | | | $\hat{\theta}_5$ | | | | |
|---|---|---|---|---|---|---|---|---|---|---|---|---|---|---|---|---|
| | | RBias (%) | SE | SD | MSE | CR (%) | RBias (%) | SE | SD | MSE | CR (%) | RBias (%) | SE | SD | MSE | CR (%) |
| 200 | | | | | | | | | | | | | | | | |
| | GS | -0.633 | 0.065 | 0.065 | 0.004 | 95.5 | -1.422 | 0.126 | 0.137 | 0.019 | 90.5 | -0.340 | 0.126 | 0.117 | 0.014 | 98.5 |
| | CC | -15.239 | 0.087 | 0.082 | 0.030 | 57.5 | -31.332 | 0.169 | 0.175 | 0.128 | 55.5 | -30.333 | 0.168 | 0.166 | 0.120 | 55.0 |
| | IPW-pooled | -9.441 | 0.110 | 0.122 | 0.024 | 81.5 | -16.927 | 0.227 | 0.312 | 0.125 | 76.0 | -15.376 | 0.224 | 0.273 | 0.098 | 78.0 |
| | PPIPW-V | -10.797 | 0.098 | 0.096 | 0.021 | 76.5 | -24.179 | 0.198 | 0.236 | 0.114 | 71.0 | -22.149 | 0.197 | 0.203 | 0.090 | 73.0 |
| | MI-naive | -4.405 | 0.081 | 0.075 | 0.007 | 93.5 | -12.129 | 0.161 | 0.142 | 0.035 | 90.5 | -11.445 | 0.161 | 0.149 | 0.035 | 88.0 |
| | MI-pooled | -3.013 | 0.079 | 0.076 | 0.007 | 94.5 | -0.078 | 0.167 | 0.168 | 0.028 | 93.5 | 1.295 | 0.165 | 0.168 | 0.028 | 95.5 |
| | PPMI-V | -2.657 | 0.074 | 0.076 | 0.006 | 94.5 | 0.633 | 0.170 | 0.167 | 0.028 | 93.5 | 1.986 | 0.168 | 0.168 | 0.029 | 96.5 |
| 1000 | | | | | | | | | | | | | | | | |
| | GS | 0.098 | 0.029 | 0.028 | 0.001 | 94.2 | 0.553 | 0.056 | 0.058 | 0.003 | 93.2 | -0.907 | 0.057 | 0.053 | 0.003 | 95.3 |
| | CC | -14.640 | 0.039 | 0.041 | 0.023 | 4.2 | -31.422 | 0.075 | 0.076 | 0.104 | 1.1 | -33.005 | 0.075 | 0.075 | 0.114 | 0.0 |
| | IPW-pooled | -5.019 | 0.070 | 0.096 | 0.012 | 76.3 | -7.326 | 0.150 | 0.254 | 0.070 | 71.6 | -11.577 | 0.151 | 0.210 | 0.057 | 72.6 |
| | PPIPW-V | -5.998 | 0.062 | 0.081 | 0.010 | 70.5 | -11.713 | 0.132 | 0.210 | 0.058 | 64.2 | -15.428 | 0.131 | 0.172 | 0.053 | 61.1 |
| | MI-naive | -2.446 | 0.035 | 0.035 | 0.002 | 88.9 | -11.166 | 0.071 | 0.063 | 0.016 | 66.3 | -12.346 | 0.071 | 0.062 | 0.019 | 61.6 |
| | MI-pooled | -0.307 | 0.034 | 0.034 | 0.001 | 95.8 | 1.365 | 0.072 | 0.074 | 0.006 | 93.7 | -0.189 | 0.073 | 0.073 | 0.005 | 95.3 |
| | PPMI-V | -0.170 | 0.033 | 0.034 | 0.001 | 93.7 | 1.216 | 0.074 | 0.074 | 0.006 | 95.8 | -0.322 | 0.074 | 0.074 | 0.005 | 95.8 |

### *Real Data Example*

We conduct an empirical study using real data from the Georgia Coverdell Acute Stroke Registry (GCASR) to evaluate our approaches. The data are collected and pooled from hospitals across the state of Georgia in the US, aimed to assess the quality of acute stroke care. For the purpose of illustration, we focus on assessing the association of four factors (i.e., Gender, Race, NIH stroke score, History of stroke) with arrival-to-CT time, which is an important quality indicator on acute stroke care. We assume that the pooled data are actually from two institutions, where the first institution has patients' demographic information (e.g., gender and race), and the second institution has clinical information (e.g., NIH stroke score and history of stroke). The outcome variable arrival-to-CT time is accessible to both institutions. We first select 31,918 patients with observations on all four independent variables and the dependent variable. The analysis of this dataset is considered a gold standard (GS). Next, we generate the missing data by artificially assigning some patients to be missing NIH stroke score through the model

$$Pr(X_3 \text{ is missing}) = \{1 + \exp(5 - Y - X_1 - X_2 - X_4)\}^{-1},$$

where $Y = \log(\text{arrival} - \text{to} - \text{CT time})$, $X_1 = 1$ if male and 0 otherwise, $X_2 = 1$ if White and 0 otherwise, $X_3 = $ NIH stroke score, $X_4 = 1$ if the patient has a history of stroke and 0 otherwise. About 45% of patients are missing an NIH stroke score according to the above criterion.

Table 3 presents parameter estimates, SEs, and p-values from the real data analyses using the aforementioned seven methods. We use $M = 20$ and $M = 100$ imputations for the MI methods (MI-pooled and PPMI-V). The results are quite close, so we only present those using $M = 20$. Consistent with our simulations, our data analysis results in Table 3 demonstrate that PPIPW-V behaves in the same way as IPW-pooled and PPMI-V perform as well as MI-pooled. Based on our analysis, it appears that there is a significant negative association between arrival-to-CT time and NIH stroke score by any of the estimates. The same finding is observed between arrival-to-CT time and race. The result from the CC analysis shows a negative effect of gender on arrival-to-CT time, while this conclusion does not hold by other methods. History of stroke is not shown to be statistically significant by IPW-pooled and PPIPW-V. In terms of the values of the estimates, MI-pooled and PPMI-V are closest to GS in general. MI-pooled and PPMI-V also provide relatively smaller SEs than other methods. The estimate of NIH stroke score using MI-naïve is only half of that using GS.

**Table 3**: Regression coefficients estimates of the Georgia stroke registry data.

| Characteristics | Method | Estimate | SE | P-value |
|---|---|---|---|---|
| Male (referent: female) | GS | 0.073 | 0.014 | <0.001 |
| | CC | -0.155 | 0.018 | <0.001 |
| | IPW-pooled | 0.106 | 0.031 | <0.001 |
| | PPIPW-V | 0.106 | 0.031 | <0.001 |
| | MI-naive | 0.050 | 0.014 | <0.001 |
| | MI-pooled | 0.069 | 0.014 | <0.001 |
| | PPMI-V | 0.068 | 0.014 | <0.001 |
| White (referent: African American) | GS | -0.159 | 0.014 | <0.001 |
| | CC | -0.353 | 0.018 | <0.001 |
| | IPW-pooled | -0.213 | 0.031 | <0.001 |
| | PPIPW-V | -0.213 | 0.031 | <0.001 |
| | MI-naive | -0.138 | 0.014 | <0.001 |
| | MI-pooled | -0.155 | 0.014 | <0.001 |
| | PPMI-V | -0.154 | 0.014 | <0.001 |
| NIH stroke score | GS | -0.032 | 0.001 | <0.001 |
| | CC | -0.024 | 0.001 | <0.001 |
| | IPW-pooled | -0.034 | 0.002 | <0.001 |
| | PPIPW-V | -0.034 | 0.002 | <0.001 |
| | MI-naive | -0.014 | 0.001 | <0.001 |
| | MI-pooled | -0.028 | 0.001 | <0.001 |
| | PPMI-V | -0.027 | 0.002 | <0.001 |
| History of stroke | GS | -0.041 | 0.016 | 0.012 |
| | CC | -0.227 | 0.019 | <0.001 |
| | IPW-pooled | -0.033 | 0.028 | 0.122 |
| | PPIPW-V | -0.033 | 0.028 | 0.122 |
| | MI-naive | -0.052 | 0.017 | 0.001 |
| | MI-pooled | -0.037 | 0.016 | 0.024 |
| | PPMI-V | -0.037 | 0.016 | 0.024 |

## Conclusion

The privacy-preserving methods developed in this paper have shown promising results for handling vertically partitioned incomplete data. Specifically, PPIPW-V models the weights for complete observations through logistic regression and solves its corresponding dual problem that utilizes summary statistics only. Then, we weight complete cases based on the estimated weights and solve an objective function of quadratic form by a derivative-free modified Powell's algorithm. The calculations within the algorithm can be linearly partitioned among institutions. The final least-squares estimate is proved to minimize the objective function. Our numerical studies demonstrate that PPIPW-V yields the same results as IPW-pooled. As has been reported in prior research, we should pay close attention to unstable weights when using IPW methods. PPIPW-V also tends to produce larger standard errors than imputation methods, which is inherited from IPW-pooled. Another privacy-preserving method that we propose is PPMI-V, which yields the same superior performance as MI-pooled compared to other native imputation methods. PPMI-V is flexible and can be extended to general missing data patterns and to the case where non-continuous variables are subject to missing values. For both PPIPW-V and PPMI-V, we also provide a privacy-preserving approach for calculating standard errors through bootstrap resampling, which enables statistical inference. However, it is possible that our distributed algorithms may still leak information through transmitting aggregated statistics, in which case we can further strengthen privacy through integrating differential privacy or secure multiparty computation. These extensions are of interest for future research. In addition, we have ongoing work on developing methods for the analysis of horizontally partitioned incomplete data for which the methods developed in this work are not applicable and we also plan to investigate more complex partition patterns that may include both vertical and horizontal partitions.

**Appendix: Technical Details**

*PPIPW-V Stage 1: Distributed logistic regression for response indicator in vertically partitioned data*

The logistic regression model for the response indicator becomes $Pr(s_i = \pm 1|\mathbf{z}) = 1/(1 + \exp(-s_i \mathbf{z}_i^T \beta))$, where $\beta \in \mathbb{R}^{(p+1) \times 1}$ is a vector of nuisance parameters to be estimated and $\mathbf{z}_i$ is the $i$-th row of $\mathbf{Z}$. The primal problem is to maximize the log-likelihood $l(\beta) = -\sum_{i=1}^{n} \log(1 + \exp(-s_i \mathbf{z}_i^T \beta)) - \lambda \beta^T \beta/2$. The penalty $\lambda \beta^T \beta/2$ is introduced to give a superior generalization performance, especially when $p$ is large. Instead of solving the primal problem, we solve the dual problem, which is represented by dual parameter $\psi \in \mathbb{R}^{n \times 1}$ as:

$$\min_{\psi} J(\psi) = \frac{1}{2\lambda} \sum_{i=1}^{n} \sum_{i'=1}^{n} \psi_i \psi_{i'} s_i s_{i'} \mathbf{z}_i^T \mathbf{z}_{i'} - \sum_{i=1}^{n} H(\psi_i), \qquad (1)$$

where $H(\psi_i) = -\psi \log \psi - (1-\psi) \log(1-\psi)$. It is easy to see that the linear kernel $\mathbf{z}_i^T \mathbf{z}_{i'}$ in Equation (1) can be linearly decomposed by institutions as $\mathbf{z}_i^T \mathbf{z}_{i'} = \sum_k (\mathbf{z}_i^{site_k})^T \mathbf{z}_{i'}^{site_k}$. Such decomposition builds the foundation of a privacy-preserving distributed logistic regression model over vertically partitioned data [16]. That is, each institution computes the dot products $(\mathbf{z}_i^{site_k})^T \mathbf{z}_{i'}^{site_k}$ and shares them to calculate $\mathbf{z}_i^T \mathbf{z}_{i'}$ of each pair of individuals. Since the dot product is a scalar, the exposure of it does not lead to the disclosure of $\mathbf{z}_i$. Newton-Raphson algorithm is applied to optimize $\psi$ via iterative procedures until convergence: $\hat{\psi}^{new} = \hat{\psi}^{old} - J''(\hat{\psi}^{old})^{-1} J'(\hat{\psi}^{old})$. With the estimated dual parameters $\hat{\psi} = (\hat{\psi}_1, \ldots, \hat{\psi}_n)^T$, we can get the estimated primal parameters $\hat{\beta} = ((\hat{\beta}^{site_1})^T, \ldots, (\hat{\beta}^{site_K})^T)^T$ by sending $\hat{\psi}$ to each institution: $\hat{\beta}^{site_k} = \lambda^{-1} \sum_{i=1}^{n} \hat{\psi}_i s_i \mathbf{z}_i^{site_k}$. Then, we can obtain the weight for individual $i$ by $\hat{w}_i = 1/\hat{p}_i =$

$1 + \exp(-\sum_{k=1}^{K} (\mathbf{z}_i^{site_k})^T \hat{\beta}^{site_k})$. Note that the dot product $(\mathbf{z}_i^{site_k})^T \hat{\beta}^{site_k}$ is calculated locally by each institution and then shared to others.

*PPIPW-V Stage 2: Distributed weighted linear regression for outcome in vertically partitioned data*

Modified Powell's algorithm:

- *Initialization*: Select an arbitrary orthogonal basis for $\mathbb{R}^{(p+1)}$: $\mathbf{d}^{(1)}, \ldots, \mathbf{d}^{(p+1)}$. Pick an arbitrary starting point $\tilde{\theta} \in \mathbb{R}^{(p+1)}$
- *Iteration*: Repeat the following steps $p + 1$ times.
    - Set $\theta = \tilde{\theta}$
    - For $j = 1, 2, \ldots, p, p+1$:
        - Let $\delta = \arg\min_\delta F(\theta + \delta \mathbf{d}^{(j)})$
        - Set $\theta = \theta + \delta \mathbf{d}^{(j)}$
    - For $j = 1, 2, \ldots, p$: Set $\mathbf{d}^{(j)} = \mathbf{d}^{(j+1)}$
    - Set $\mathbf{d}^{(p+1)} = \theta - \tilde{\theta}$
        - $\delta = \arg\min_\delta F(\theta + \delta \mathbf{d}^{(p+1)})$
        - Set $\tilde{\theta} = \theta + \delta \mathbf{d}^{(p+1)}$

Of note, for the objective function of sums of weighted errors, given any direction $\mathbf{d}$,

$$\delta = \arg\min_\delta F(\theta + \delta \mathbf{d}) = \frac{(Y - \mathbf{X}\theta)^T \mathbf{V}(\mathbf{X}\mathbf{d})}{(\mathbf{X}\mathbf{d})^T \mathbf{V}(\mathbf{X}\mathbf{d})} = \frac{\gamma^T \mathbf{V} \eta}{\eta^T \mathbf{V} \eta},$$

where $\gamma = Y - \mathbf{X}\theta$ and $\eta = \mathbf{X}\mathbf{d}$. Similar to the data are vertically partitioned (i.e., $\mathbf{X} = (\mathbf{X}^{site_1}, \ldots, \mathbf{X}^{site_K})$), we partition the direction $\mathbf{d}$ and the vector of parameters $\theta$ as $\mathbf{d} = ((\mathbf{d}^{site_1})^T, \ldots, (\mathbf{d}^{site_K})^T)^T$ and $\theta = ((\theta^{site_1})^T, \ldots, (\theta^{site_K})^T)^T$ accordingly. Therefore, $\gamma = Y - \sum_{k=1}^{K} \mathbf{X}^{site_k} \theta^{site_k}$ and $\eta = \sum_{k=1}^{K} \mathbf{X}^{site_k} \mathbf{d}^{site_k}$. Such linear decompositions allow us to obtain $\delta$ by only sharing the locally calculated summary statistics (i.e., $\mathbf{X}^{site_k} \theta^{site_k}$, $\mathbf{X}^{site_k} \mathbf{d}^{site_k}$).

## Acknowledgments

QL was partly supported by the National Institutes of Health grant R01GM124111. XJ is CPRIT Scholar in Cancer Research (RR180012), and he was supported in part by Christopher Sarofim Family Professorship, UT Stars award, UTHealth startup, the National Institute of Health (NIH) under award number R01AG066749, R01GM114612, R01GM124111, and U01TR002062, and the National Science Foundation (NSF) RAPID #2027790.